\documentclass{cernrep}
\begin{document}
\title{Prompt photon production with POWHEG}
\author{M. Klasen %$^a$, C. Klein-B\"osing$^b$, H. Poppenborg$^b$
%\thanks{On leave from another institue somewhere.}
}
\institute{% $^a$
 Institut f\"ur Theoretische Physik, Westf\"alische Wilhelms-Universit\"at
 M\"unster, Wilhelm-Klemm-Stra\ss{}e 9, D-48149 M\"unster, Germany
% \\ $^b$ Institut f\"ur Kernphysik, Westf\"alische Wilhelms-Universit\"at M\"unster, Wilhelm-Klemm-Stra\ss{}e 9, D-48149 M\"unster, Germany
}

\begin{abstract}
%Each paper should be preceded by a short abstract of not more 
%than 150~words, which should be written as a single paragraph 
%and should not contain references and notes.
We present a calculation of prompt photon and associated photon-jet
production at next-to-leading order that is consistently matched to
parton showers with POWHEG. Specific issues that appear in photon
radiation are discussed. Numerical results are compared to pp
collision data at RHIC energies and shown to describe the data better
than inclusive next-to-leading order calculations or those with
leading order Monte Carlo generators like PYTHIA alone.
\end{abstract}


\maketitle

\section{Introduction}
\label{sec:1}

\vspace*{-105mm}
MS-TP-17-10\\
\vspace*{95mm}

\noindent Theoretical calculations of prompt photon production at hadron
colliders have a long-standing tradition. Their importance derived
originally from understanding perturbative QCD \cite{Owens:1986mp}, e.g.\
the fractional quark charges or renormalisation group effects, whereas
photon pair production is currently an important background in one of the
Higgs-boson discovery channels at the LHC \cite{Aad:2014eha}. In heavy-ion
collisions, thermal photons are an important probe to determine
the effective temperature of the created Quark-Gluon Plasma (QGP)
through their characteristic exponential transverse momentum ($p_T^\gamma$)
spectrum \cite{Adam:2015lda}.

Leading-order (LO) calculations, supplemented by parton showers (PS)
and hadron fragmentation and implemented in Monte Carlo generators
like PYTHIA 8 \cite{Sjostrand:2007gs}, provide detailed information
on the final state and are indispensable tools in the experimental
analyses. Inclusive next-to-leading order (NLO) calculations like
JETPHOX \cite{Aurenche:2006vj} employ, in contrast, inclusive
fragmentation functions (FFs) like BFG II \cite{Bourhis:1997yu} and
have a smaller theoretical scale uncertainty. With NLO+PS Monte Carlo
methods like POWHEG \cite{Frixione:2007vw} it is possible to
combine the advantages of both approaches.
We have recently applied this method to prompt photon and associated
photon-jet production \cite{Jezo:2016ypn}. We review the theoretical
approach in Sec.\ \ref{sec:2} and then demonstrate its phenomenological
advantages by applying it to PHENIX data from RHIC \cite{Adare:2012vn}
in Sec.\ \ref{sec:3}. Our conclusions are given in Sec.\ \ref{sec:4}.

\section{Theoretical approach}
\label{sec:2}

The POWHEG method requires first the recalculation of Born processes
with spin and colour correlations, in this case for $q\bar{q}\to
\gamma g$ and the QCD Compton process $qg\to\gamma q$. Next, the
virtual corrections must be recomputed and their ultraviolet and
infrared divergences consistently renormalised and subtracted,
respectively. The real emission amplitudes must not be subtracted,
since POWHEG does so automatically. While the hardest radiation is
thus generated first, subsequent emissions are produced by the
parton shower implemented, e.g., in PYTHIA \cite{Sjostrand:2007gs},
leading always to detailed events with positive weights.
This method applies not only to QCD radiation, but can be generalised
to QED radiation. One must then check that fragmentation photons
like those produced in $e^+e^-$ collisions at LEP are as well
described with PSs as with inclusive FFs \cite{Hoeche:2009xc}.

There are, however, a number of specific issues for photons.
First, QED radiation is suppressed with respect to QCD radiation
by the smaller coupling, colour factors and multiplicities, so
that it must be artificially enhanced. Second, the hard scale is
not necessarily the $p_T^\gamma$ of the observed photon, but may be
the $p_T$ of an underlying QCD parton. Third, the radiation process
$q\to q\gamma$ must be correctly symmetrised, i.e.\ also include
the process $q\to\gamma q$. Finally, inclusive photon production
$pp\to\gamma+X$ has a collinear divergence at low $p_T^\gamma$, which
must be carefully regularised so that sensitivity to the low-$p_T^\gamma$
region important for heavy-ion collisions and independence on the
regularisation method are simultaneously maintained.

\section{Prompt photons and photon-jet correlations at RHIC}
\label{sec:3}

We now compare the implementation of prompt photons in POWHEG to
data obtained by the PHENIX collaboration in pp collisions at RHIC
\cite{Adare:2012vn}. These data represent an important baseline
for studies of the QGP produced in heavy-ion collisions and should
be described by prompt (direct and fragmentation) contributions
alone.

\begin{figure}[!t]
 \centering
 \includegraphics[width=0.91\linewidth]{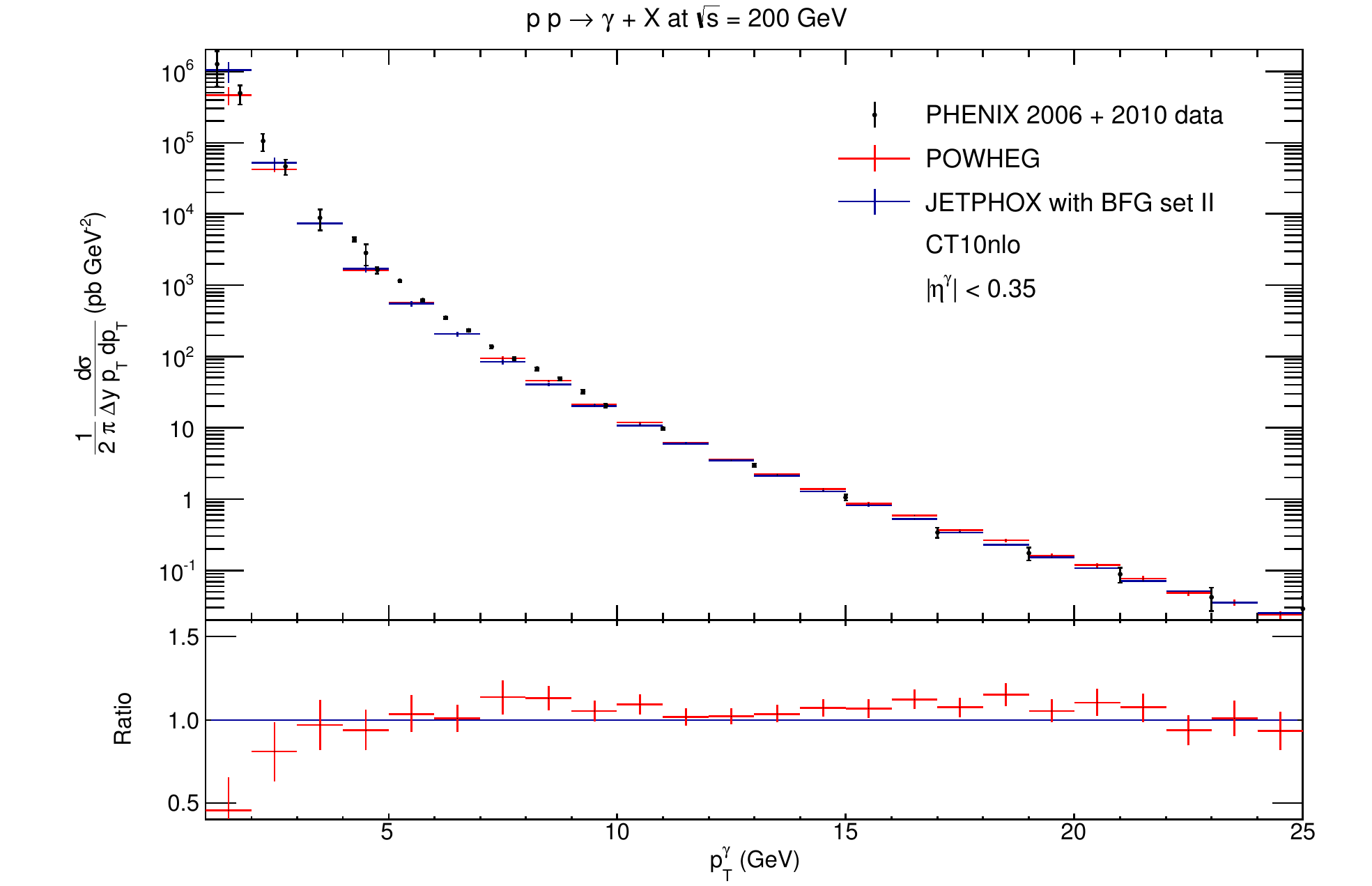}
 \caption{\label{fig:01}Inclusive photon production in pp collisions at RHIC
 with a centre-of-mass energy of 200 GeV. PHENIX data (black) are compared with
 NLO+PS predictions with POWHEG+PYTHIA (red) and NLO predictions with JETPHOX
 (blue).}
\end{figure}

In Fig.\ \ref{fig:01} the inclusive photon distribution in
transverse momentum is shown at the RHIC centre-of-mass energy
of $\sqrt{s}=200$ GeV. As expected for this inclusive quantity,
both the NLO calculation with JETPHOX (blue) and the NLO+PS
calculation with POWHEG+PYTHIA (red) describe the data up to
25 GeV. The two theoretical predictions differ only at very low 
$p_T^\gamma$, where fragmentation contributions dominate. There,
the POWHEG+PYTHIA predictions are lower than those with JETPHOX
and seem to agree better with the PHENIX data, although both
are consistent with the data within error bars.

A quantity that is more sensitive to the treatment of the photon
fragmentation process is the fraction of isolated photons, defined
by a hadronic energy fraction of less than 10\% of the energy of
the photon in a cone of radius $R=\sqrt{(\Delta\eta)^2+
(\Delta\phi)^2}\leq0.5$ around it. The comparison with JETPHOX
in Fig.\ 13 of the experimental publication led to the conclusion
that neither BFG II nor GRV \cite{Gluck:1992zx} photon FFs
described this fraction correctly, even after accounting for
statistical, systematic, and theoretical scale uncertainties
\cite{Adare:2012vn}.

\begin{figure}[!t]
 \centering
 \includegraphics[width=0.91\linewidth]{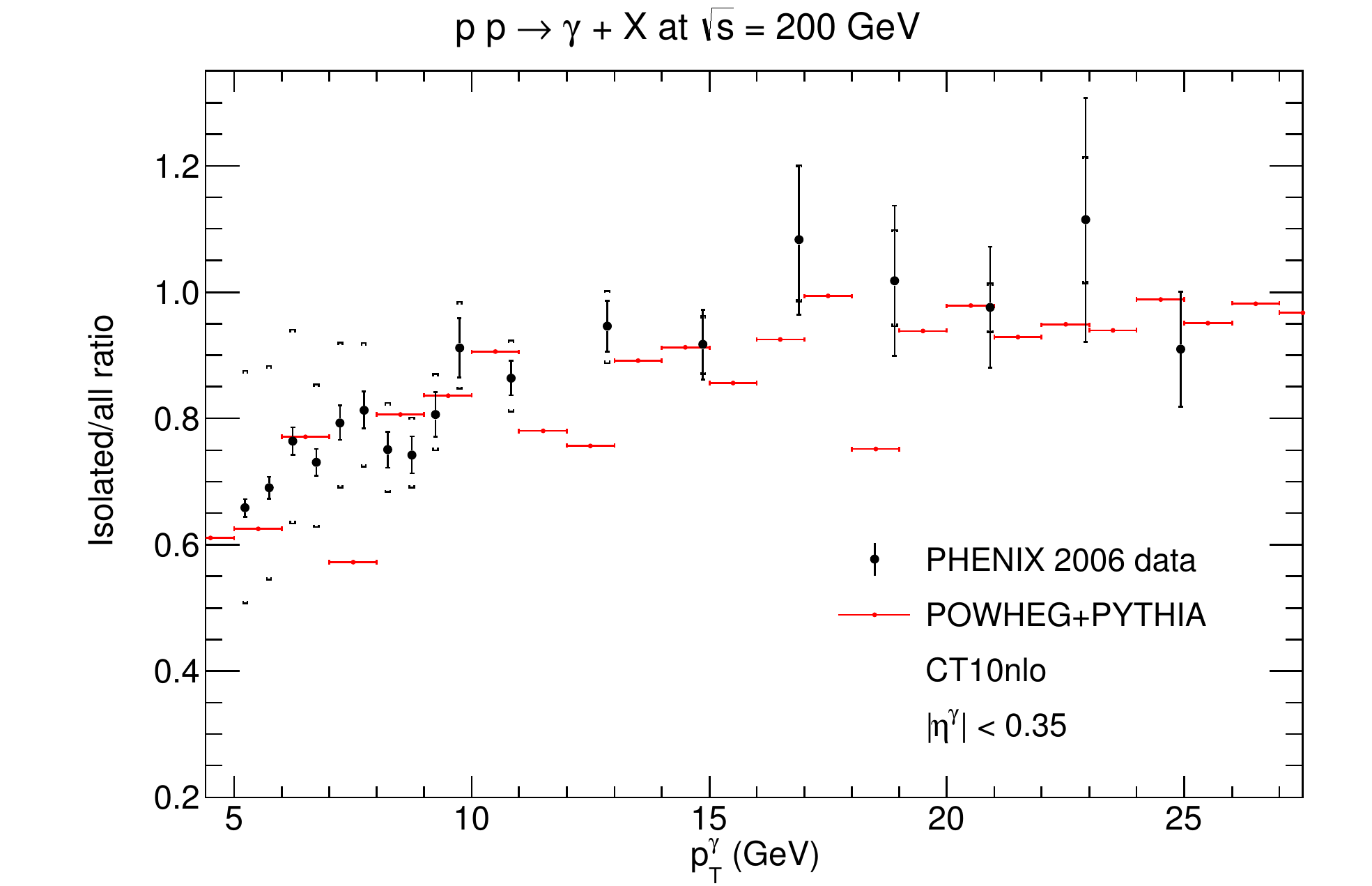}
 \caption{\label{fig:02}Isolated photon production in pp collisions at RHIC
 with a centre-of-mass energy of 200 GeV. PHENIX data (black) are compared with
 NLO+PS predictions with POWHEG+PYTHIA (red).}
\end{figure}

As one can see in Fig.\ \ref{fig:02}, the fraction of isolated
photons predicted by POWHEG+PYTHIA describes the data very well,
apart from a few fluctuations due to limited Monte Carlo statistics.
The fraction rises from about 60\% at low $p_T^\gamma$, where a
substantial fraction of photons does not survive the isolation
cut due to collinear parton radiation, to almost unity for intermediate
and large $p_T^\gamma$. There, photons are rather produced back-to-back
with a recoiling jet and have little near-side hadronic energy.

\begin{figure}[!h]
 \centering
 \includegraphics[width=0.91\linewidth]{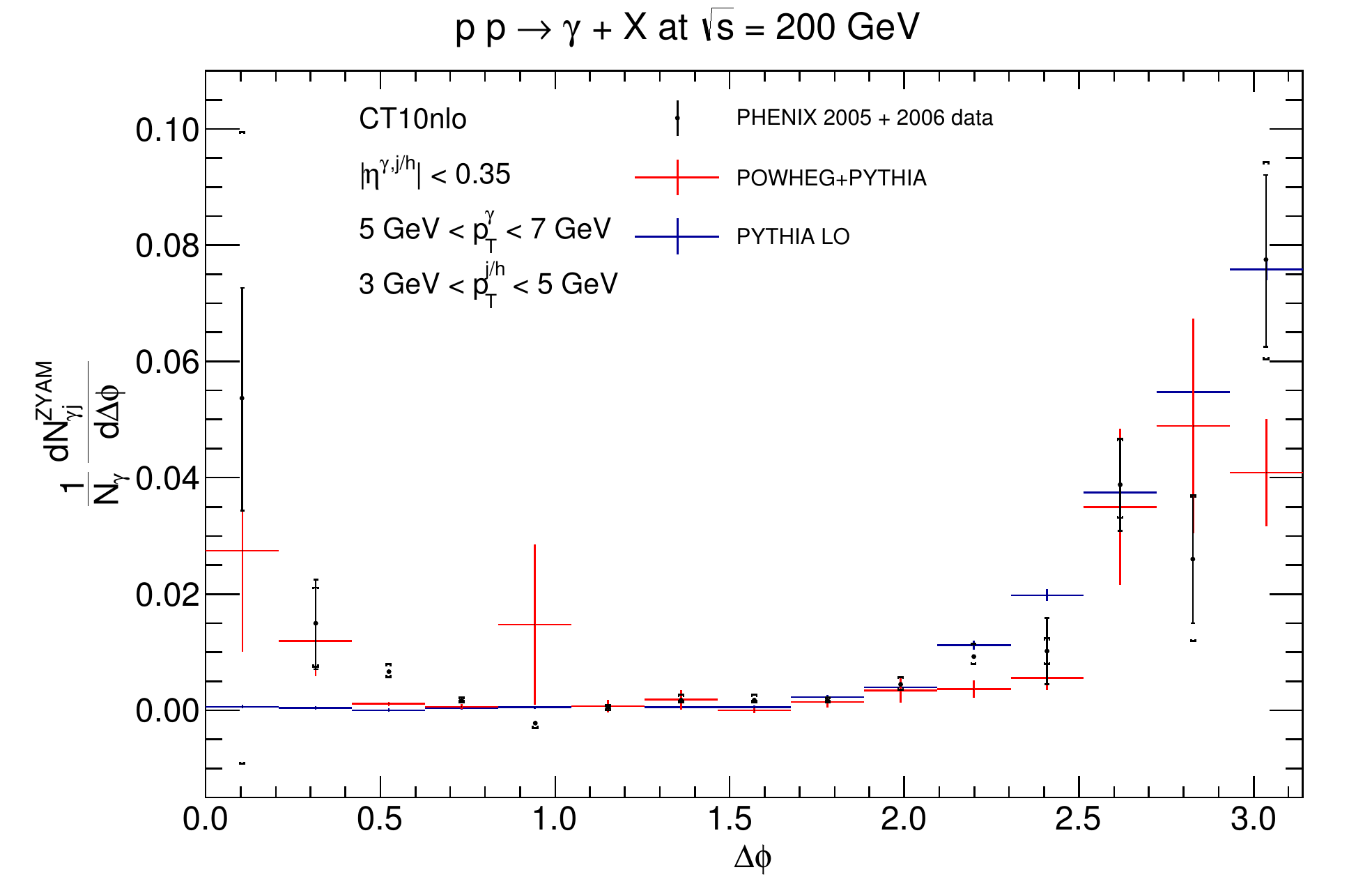}
 \caption{\label{fig:03}Associated photon+jet production in pp collisions at RHIC
 with a centre-of-mass energy of 200 GeV. PHENIX data (black) are compared with
 NLO+PS predictions with POWHEG+PYTHIA (red) and LO+PS predictions with PYTHIA
 (blue).}
\end{figure}

When both the photon and the recoiling jet are experimentally
measured, the two may be correlated either in transverse momentum
or in azimuthal angle $\Delta\phi$. Both quantities are sensitive
to modifications of the hadronic jet due to rescatterings in the
QGP, but also to higher-order perturbative QCD effects
\cite{Klasen:1995xe}. Fig.\ \ref{fig:03} shows the distribution
in azimuthal angle, subtracted for decay photons assuming a
Zero-Yield at Minimum (ZYAM) and normalised to the total number
of trigger photons. The near- ($\Delta\phi\sim0$) and away-side
($\Delta\phi\sim\pi$) regions are clearly visible, but only the
latter is correctly described by PYTHIA alone, while both are
reproduced with POWHEG+PYTHIA. As expected and as we have shown
in our original paper \cite{Jezo:2016ypn}, the near-side region
is dominated by fragmentation photons, which requires a proper
matching of NLO and PS contributions.
 
\section{Conclusion}
\label{sec:4}

To conclude, we have reviewed our recent implementation of prompt photon
production in POWHEG. Our calculations now allow for predictions of
prompt photon and photon-jet associated production at hadron colliders
with reduced theoretical scale uncertainties and sufficient detail of
the produced final state. We have successfully applied our calculations
to PHENIX data from RHIC. Applications to the LHC will be presented in
a forthcoming publication.

\section*{Acknowledgements}

We thank T.\ Jezo and F.\ K\"onig for their collaboration and the organisers of the
conference for the kind invitation.
This work has been supported by the BMBF under contract 05H15PMCCA and by the  DFG
through the Research Training Network 2149 ``Strong and weak interactions -
from hadrons to dark matter''.

\end{document}